# Vision Based Game Development Using Human Computer Interaction

Ms.S.Sumathi  
Bharath University  
Chennai,India

Dr.S.K.Srivatsa  
St.Joseph College of Engineering  
Chennai,India

Dr.M.Uma Maheswari  
Bharath University  
Chennai,India

*Abstract*— **A Human Computer Interface (HCI) System for playing games is designed here for more natural communication with the machines. The system presented here is a vision-based system for detection of long voluntary eye blinks and interpretation of blink patterns for communication between man and machine. This system replaces the mouse with the human face as a new way to interact with the computer. Facial features (nose tip and eyes) are detected and tracked in real-time to use their actions as mouse events. The coordinates and movement of the nose tip in the live video feed are translated to become the coordinates and movement of the mouse pointer on the application. The left/right eye blinks fire left/right mouse click events. The system works with inexpensive USB cameras and runs at a frame rate of 30 frames per second.**

## I. INTRODUCTION

One of the promising fields in artificial intelligence is HCI. Human-Computer Interface (HCI) can be described as the point of communication between the human and a computer. HCI aims to use human features to interact with the computer. The system tracks the computer user's movements with a video camera and translates them into the movements of the mouse pointer on the screen. The tip of the user's nose can be tracked and captured with a webcam and monitor its movements in order to translate it to some events that communicate with the computer. In our system, the nose tip is the pointing device, because of the location and shape of the nose, as it is located in the middle of the face it is more comfortable to use it as the feature that moves the mouse pointer and defines its coordinates, and it is located on the axis that the face rotates about, so it basically does not change its distinctive convex shape which makes it easier to track as the face moves. We use the human feature Eyes to simulate mouse clicks, so the user can fire their events as he blinks. While different devices were used in HCI (e.g. infrared cameras, sensors, microphones) we used a webcam that affords a moderate resolution and frame rate as the capturing device in order to make the ability of using the program affordable for all individuals.

We present an algorithm that distinguishes true eye blinks from involuntary ones, detects and tracks the desired facial features precisely, and fast enough to be applied in real-time. In our work we are trying to implement this in playing video games. Getting the player physically involved in the game provides a more immersive experience and a feeling of taking direct part rather than just playing as an external beholder. Already motion sensors have been implemented to recognize physical activity, we can even use finger gestures in playing games but we increase human computer interaction by using our eyes in playing games.

We propose an accurate algorithm that distinguishes true eye blinks from involuntary ones, detects and measures their duration, and fast enough to be applied in real-time to control a non –intrusive interface for computer users in playing games. This system can be used in playing games like first shooter game, a role playing game and an action game. When we compare eye-based and mouse-based control it is found that using an eye tracker can increase the immersion and leads to a stronger feeling of being part of the game. We are tracking the nose movements and eye blinks for playing the games, the nose movements are interfaced with the mouse movements and the eye blinks are interfaced with the mouse clicks. The eye-movement or eye blink controlled human-computer interface systems are very useful for persons who cannot speak or use hands to communicate. There are no lighting requirements or offline templates needed for the proper functioning of the system. It works with inexpensive USB cameras and runs at a frame rate of 30 frames per second.

The automatic initialization phase is triggered by the analysis of the involuntary blinking of the current computer user, which creates an online template of the eye to be used for tracking. This phase occurs each time the current correlation score of the tracked eye falls below a defined threshold in order to allow the system to recover and regain its accuracy in detecting the blinks. This system can be utilized by users for applications that require mouse clicks as input for e.g. games. The main contribution of this paper is to provide a reimplementation of this system that is able to run in real time at 30 frames per second on readily available and affordable webcams in playing video games.

## II. RELATED WORK

With the growth of attention about computer vision, the interest in HCI has increased proportionally. Different human features and monitoring devices were used to achieve HCI, but during our research we were only into works that involved the use of facial features and webcams.

The current evolution of computer technologies has enhanced various applications in human-computer interface. Face and gesture recognition is a part of this field, which can be applied in various applications such as in robotic, security system, drivers monitor, and video coding system.

We noticed a large diversity of the facial features that were selected, the way they were detected and tracked, and the functionality that they presented for the HCI. Researchers chose different facial features: eye pupils, eyebrows, nose tip, lips, eye lids' corners, mouth corners for each of which they provided an explanation to choose that particular one.

Different detection techniques were applied where the goal was to achieve more accurate results with less processing time. To control the mouse pointer various points were tracked ranging from the middle distance between the eyes, the middle distance between the eyebrows, to the nose tip. To simulate mouse clicks; eye blinks, mouth opening/closing, and sometimes eyebrow movement were used. Each HCI method that we read about had some drawbacks, some methods used expensive equipments, some were not fast enough to achieve real-time execution, and others were not robust and precise enough to replace the mouse. We tried to profit from the experience that other researchers gained in the HCI field and added our own ideas to produce an application that is fast, robust, and useable.

Eye movement events detected in EOG signals such as saccades, fixations and blinks have been used to control robots or a wearable system for medical care givers. Patmore et al. described a system that provides a pointing device for people with physical disabilities. All of these systems use basic eye movements or eye-gaze direction but they do not implement movement sequences which provide a more versatile input modality for gaming applications.

## III. SYSTEM OVERVIEW

This system design can be broken down into three modules, (1) Facial Features tracking (2) Integrating Nose tip movements with the mouse cursor (3) Replacing the eye blinks with the mouse click events.

**Face Detection**

In this module, we propose a real-time face detection algorithm using Six-Segmented Rectangular (SSR) filter, distance information, and template matching technique. Since human face is a dynamic object and has a high degree of variability, we propose the method combine both feature-based and image-based approach to detect the point between the eyes by using Six-Segmented Rectangular filter (SSR filter). The proposed SSR filter, which is the rectangle divided into 6 segments, operates by using the concept of bright-dark relation around Between-the-Eyes area. Between-the-Eyes is selected as face representative in our detection because its characteristic is common to most people and is easily seen for a wide range of face orientation.

Firstly, we scan a certain size of rectangle divided into six segments throughout the face image. Then their bright-dark relations are tested if its center can be a candidate of Between-the-Eyes. Next, the distance information obtained from stereo camera and template matching is applied to detect the true Between-the-Eyes among candidates. Between-the-Eyes has dark part (eyes and eyebrows) on both sides, and has comparably bright part on upper side (forehead), and lower side (nose and cheekbone). This characteristic is stable for any facial expression.

We use an intermediate representation of image called integral image to calculate sums of pixel values in each segment of SSR filter. Firstly, SSR filter is scanned on the image and the average gray level of each segment is calculated from integral image. Then, the bright-dark relations between each segment are tested to see whether its center can be a candidate point for Between-the- Eyes. Next, the stereo camera is used to find the distance information and the suitable Between-the- Eyes template size.. Finally the true Between-the-Eyes can be detected.

*1) Integral Image*

The SSR filter is computed by using intermediate representation for image called integral image. For the original image i(x, y), the integral image is defined as,

$$ii(x, y) = \sum_{x' \leq x} \sum_{y' \leq y} i(x', y') \quad (1)$$

The integral image can be computed in one pass over the original image by the following pair of recurrences.

$$s(x, y) = s(x, y-1) + i(x, y) \quad (2)$$
$$ii(x, y) = ii(x-1, y) + s(x, y) \quad (3)$$

Where *s(x, y)* is the cumulative row sum,

*s(x, -1) = 0*, and *ii(-1, y) = 0*.

Using the integral image, the sum of pixels within rectangle D ($r_s$) can be computed at high speed with four array references as shown in Fig.1.

$$s_r = (ii(x, y) + ii(x - W, y - L)) - (ii(x - W, y) + ii(x, y - L)) \quad (4)$$

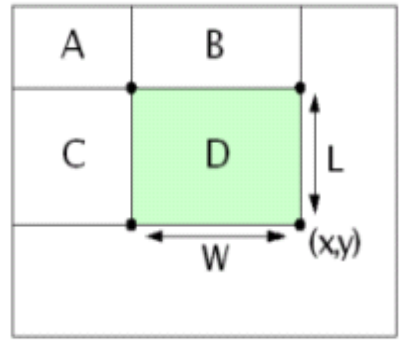

*Figure 1. Integral Image*

2) SSR filter

At the beginning, a rectangle is scanned throughout the input image. This rectangle is segmented into six segments as shown in Fig.2 (a).

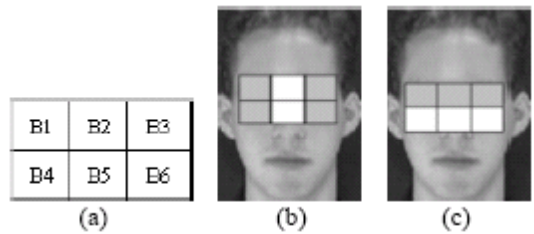

*Figure 2. SSR Filter*

We denote the total sum of pixel value of each segment (B1 B6) as 1 6 b b S S. The proposed SSR filter is used to detect the Between-the-Eyes based on two characteristics of face geometry.

(1) The nose area ($S_n$) is brighter than the right and left eye area ($S_{er}$ and $S_{el}$, respectively) as shown in Fig.2 (b), where

$$S_n = S_{b2} + S_{b5}$$
$$S_{er} = S_{b1} + S_{b4}$$
$$S_{el} = S_{b3} + S_{b6}$$

Then,

$$S_n > S_{er} \quad (5)$$
$$S_n > S_{el} \quad (6)$$

(2) The eye area (both eyes and eyebrows) ($S_e$) is relatively darker than the cheekbone area (including nose) ($S_c$) as shown in Fig. 2 (c), where

$$S_e = S_{b1} + S_{b2} + S_{b3}$$
$$S_c = S_{b4} + S_{b5} + S_{b6}$$

Then,

$$S_e < S_c \quad (7)$$

When expression (5), (6), and (7) are all satisfied, the center of the rectangle can be a candidate for Between-the-Eyes.

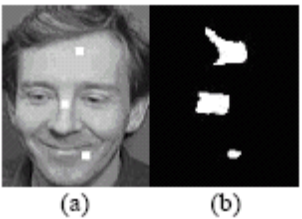

*Figure 3. Between-the-Eyes candidates from SSR filter*

In Fig.3 (b), the Between-the-Eyes candidate area is displayed as the white areas and the non-candidate area is displayed as the black part. By performing labeling process on Fig. 3 (b), the result of using SSR filter to detect Between-the-Eyes candidates is shown.

Because the SSR filter extracts not only the true Between-the- Eyes but also some false candidates, so we use the average Between-the-Eyes template matching technique to solve this problem.

To evaluate the candidates, we define the Between the- Eyes pattern as $p_{mn}$ (m=0,...,31, n = 0, ...., 15).

Then right and left half of $p_{mn}$ is re-defined again separately as $p^r_{ij}$ (i=0,...,15, j = 3, ...., 15) and $p^l_{ij}$ (i=0,...,15, j = 3, ...., 15), respectively, each has been converted to have average value of 128 and standard deviation of 64.

Then the left mismatching value ($D_l$) and the right mismatching value ($D_r$) are calculated by using the following equation.

$$D_l = \sum \frac{(p_{ij}^l - t_{ij}^l)^2}{v_{ij}^l} \quad (8)$$

$$D_r = \sum \frac{(p_{ij}^r - t_{ij}^r)^2}{v_{ij}^r} \quad (9)$$

The processing flow of Real-Time face detection system is shown in Fig. 4.

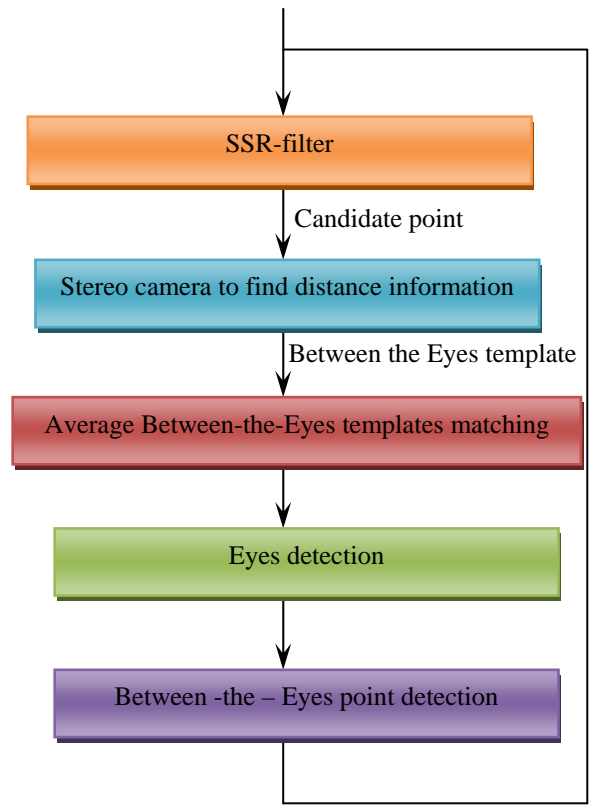

*Figure 4. Processing Flow of Real-Time Face Detection*

After extracting the templates, we pass them to the support vector machine in order to classify them. Positive classification results mean true faces, while negative ones mean false faces. Since the program will be used by one person at a time, we need to pick one of the positive results as the final detected face. To achieve that, we pick the highest positive result, but before doing so, we will multiply each positive result by the area of the cluster that its template represents.

**Finding the Nose Tip**

After locating the eyes, the final step is to find the nose tip. From figure 5 we can see that the blue line defines a perfect square of the pupils and outside corners of the mouth; the nose tip should fall inside this square, so this square becomes our region of interest in finding the nose tip.

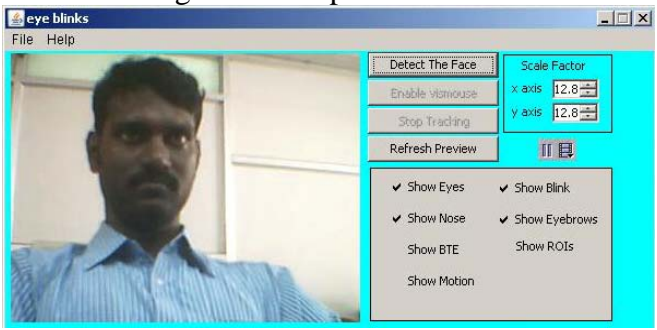

*Figure 5 The square that forms the ROI*

The nose tip has a convex shape so it collects more light than other features in the ROI because it is closer to the light source. In horizontal intensity profiles we add vertically to each line the values of the lines that precedes it in the ROI, so since that the nose bridge is brighter than the surrounding features the values should accumulate faster at the bridge location; in other words the maximum value of the horizontal profile gives us the 'x' coordinate of the nose tip.

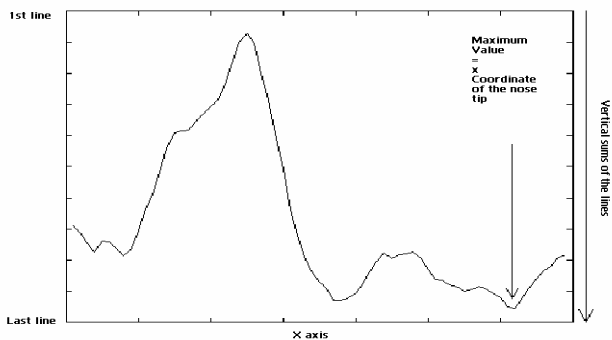

*Figure 6:The horizontal profile of ROI*

In vertical intensity profiles we add horizontally to each column the values of the columns that precedes it in the ROI the same as in the horizontal profile, the values accumulate faster at the nose tip position so the maximum value gives us the 'y' coordinate of the nose tip. From both, the horizontal and vertical profiles we were able to locate the nose tip position.

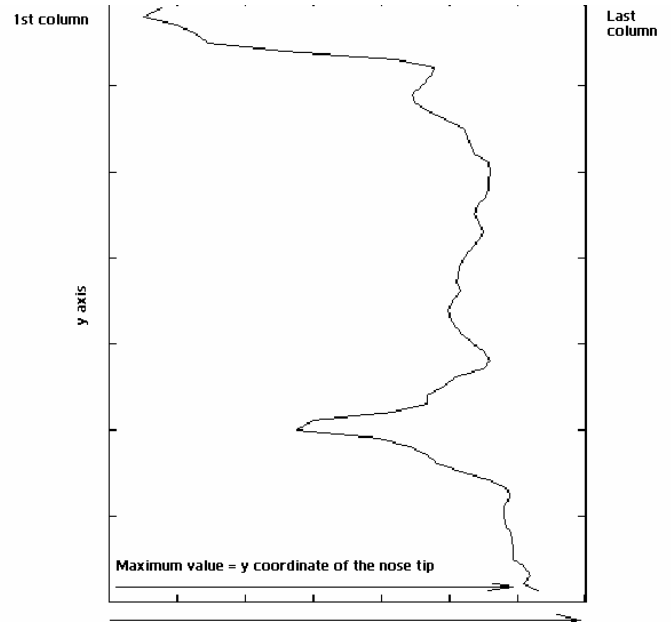

*Figure 7: Horizontal Sums of the columns*

After locating the nose bridge we need to find the nose tip on that bridge. Since each NBP represents the brightest S2 sector on the line it belongs to, and that S2 sector contains the accumulated vertical sum of the intensities in that sector from the first line to the line it belongs to, we will be using this information to locate the nose tip. Nose trills are dark areas, and the portion that they add to the accumulated sum in the horizontal profile is smaller than the contribution of other areas; in other words each NBP will add with its S2 sector a certain amount to the accumulated sum in the horizontal profile, but the NBP at the nose trills location will add a smaller amount, we will notice a local minima at the nose trills location, by locating this local minima we take the NBP that corresponds to it as the nose trills location, and the next step is to look for the nose tip above the nose trills. Since the nose tip is brighter than other features it will donate with its S2 sector to the accumulated sum more than other NBPs, which means a local maxima in the first derivate, so the location of the nose tip is the location of the NBP that corresponds to the local maxima that is above the local minima in the first derivate. Tracking the nose tip will be achieved by template matching inside the ROI.

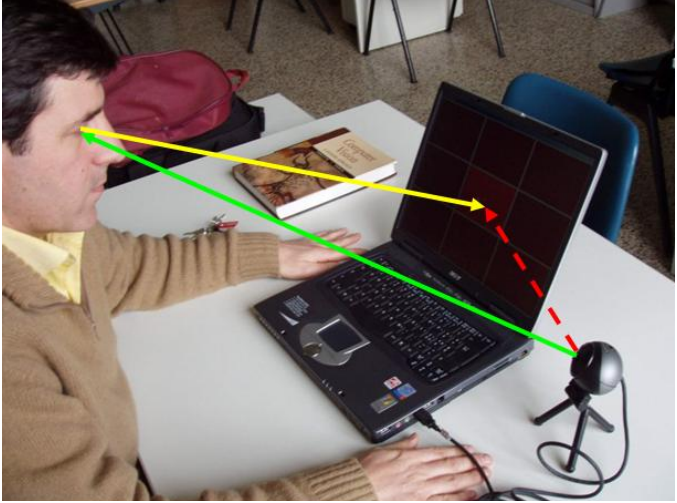

*Figure 8:The final result of the face and eye detection process*

### Hough Transform

The Hough transform is a technique which can be used to isolate features of a particular shape within an image. In our proposal we use this to find our eye brows. The classical Hough transform is most commonly used for the detection of regular curves such as lines, circles, ellipses, etc.

The Hough transform can be used to identify the parameter of a curve which best fits a set of given edge points. This edge description is commonly obtained from a feature detecting operator such as the Roberts Cross, Sobel or Canny edge detector and may be noisy, i.e. it may contain multiple edge fragments corresponding to a single whole feature. Furthermore, as the output of an edge detector defines only where features are in an image, the work of the Hough transform is to determine both what the features are (i.e. to detect the feature(s) for which it has a parametric (or other) description) and how many of them exist in the image. To find the eyebrow line from the set of thresholding points we apply the Hough transform. Sometimes Hough transform gives several lines so we approximate them to a final line which is the eye brow.

### Motion Detection

To detect motion in a certain region we subtract the pixels in that region from the same pixels of the previous frame, and at a given location (x,y); if the absolute value of the subtraction was larger than a certain threshold, we consider a motion at that pixel.

### Blink Detection

We apply blink detection in the eye's ROI before finding the eye's new exact location. The blink detection process is run only if the eye is not moving, because when a person uses the mouse and wants to click, he moves the pointer to the desired location, stops, and then clicks, so basically the same for using the face, the user moves the pointer with the tip of the nose, stops, then blinks. To detect a blink we apply motion detection in the eye's ROI; if the number of motion pixels in the ROI is larger than a certain threshold we consider that a blink was detected, because if the eye is still, and we are detecting a motion in the eye's ROI, that means that the eyelid is moving which means a blink. In order to avoid multiple blinks detection while they are a single blink (because motion pixels will appear while the eye is closing and reopening), the user can set the blink's length, so all blinks which are detected in the period of the first detected blink are omitted.

### IV Conclusion

The proposed system is the best system for the users to play games interactively. The automatic initialization phase is greatly simplified in this system, with no loss of accuracy in locating the user's eyes and choosing a suitable open eye template. Another improvement in this system is, it is compatible with inexpensive USB cameras, as opposed to the high- resolution cameras. The experiments indicate that the system performs equally well in extreme lighting conditions. The accuracy percentages in all the cases were approximately the same as those that were retrieved in normal lighting conditions.

Another important consideration is the placement and orientation of camera. The experiments showed that placing the camera below the user's head resulted in desirable functioning of the system. However, if the camera is placed too high above the user's head, in such a way that it is aiming down at the user at a significant angle, the blink detection is no longer as accurate. This is caused by the very small amount of variation in correlation scores as the user blinks, since nearly all that is visible to the camera is the eyelid of the user. Thus, when positioning the camera, it is beneficial to the detection accuracy to maximize the degree of variation between the open and closed eye images of the user. Higher frame rates and finer camera resolutions could lead to more robust eye detection that is less restrictive on the user, while increased processing power could be used to enhance the tracking algorithm to more accurately follow the user's eye and recover more gracefully when it is lost.

AUTHORS PROFILE

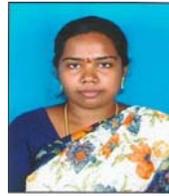

She has completed her Bachelor of engineering in Computer Science from Raja Rajeswari engineering College, Anna University and Master of Engineering from P.S.N.A Engineering college, Anna University. She is currently doing her Ph.D in Bharath University. Her research interest includes image processing, computer vision and database management systems.

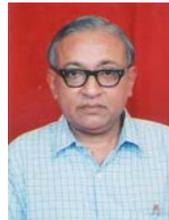

Dr.S.K.Srivatsa was born at Bangalore on 21st July 1945. He received his Bachelor of Electronics and Telecommunication Engineering degree (Honors) from Jadavpur University (securing first rank and two medals). Master degree in Electrical Commuication Engineering (With distinction) from Indian Institute of Science and Ph.D also from Indian Institute of Scienc, Bangalore. In July 2005, he retired as professor of Electronics engineering from Anna University. He has taught twenty-two different Courses at P.G level during the last 33 years. He has functioned as a Member of the Board of studies in some Educational Institutions. His name is included in the Computer Society of India database of Resource professionals. He has received about a dozen awards. He has produced 24 Ph.Ds. He is the author of well over 400 publications.

.

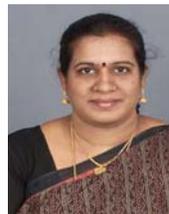

Dr.M.Uma maheswari received her bachelor of science in Computer Science from Bharathidasan university in 1995, Master of Computer Applications in Computer Sience from Bharathidasan University in1998, M.Phil in Computer Science from Alagappa University, Karaikudi in 2005, Master of Technology in Computer Science from Mahatma Gandhi Kasi Vidyapeeth university in 2005 and Ph.D in Computer Science from Magadh Universty, Bodh Gaya in 2007. She has 10 years of teahing experience and guided 150 M.C.A projects,23 M.Tech projects and 6 Ph.D research works.